\documentclass[twocolumn,prl,superscriptaddress,showpacs]{revtex4}
\usepackage[colorlinks,bookmarks=false,citecolor=blue,linkcolor=red,urlcolor=blue]{hyperref}
\usepackage{graphicx,epstopdf,color}
\usepackage{amsfonts}
\usepackage{amsmath,amssymb,mathrsfs}
\usepackage{bm}

\newcommand{\bs}[1]{\boldsymbol{#1}}

\newcommand{\ket}[1]{\left|#1\right\rangle}

\renewcommand{\i}{\text{i}}

\newcommand{\z}{\text{z}}

\newcommand{\dw}{\downarrow}

\newcommand{\vac}{\left|0\right\rangle}

\renewcommand{\a}{\alpha}
\renewcommand{\b}{\beta}
\renewcommand{\c}{\gamma}
\newcommand {\ea}{\eta_{\alpha}}
\newcommand {\eb}{\eta_{\beta}}
\newcommand {\ec}{\eta_{\gamma}}

\newcommand {\eab}{\bar\eta_{\alpha}}
\newcommand {\ebb}{\bar\eta_{\beta}}

\newcommand {\Sa}{S_{\alpha}}
\newcommand {\Sb}{S_{\beta}}
\newcommand {\Sc}{S_{\gamma}}

\newcommand {\bSa}{\bs{S}_{\alpha}}
\newcommand {\bSb}{\bs{S}_{\beta}}
\newcommand {\bSc}{\bs{S}_{\gamma}}

\newcommand {\OaS}{\Omega_{\alpha}^{s}}

\def\s{\scriptscriptstyle}

\newlength{\ytlength}
\def\ie{{i.e.},\ }

\allowdisplaybreaks[1]
\begin{document}
\title{Parent Hamiltonian for the non-Abelian chiral spin liquid}
\author{Martin Greiter}
\affiliation{Institut f\"ur Festsk\"orperphysik, Postfach 3640, KIT, D 76021
  Karlsruhe, Germany}
\author{Darrell F. Schroeter}
\affiliation{Department of Physics, Reed College, Portland, OR 97202, USA}
\author{Ronny Thomale}
\affiliation{Department of Physics, Stanford University, Stanford, CA 94305, USA}

\pagestyle{plain}

\begin{abstract}
  We construct a parent Hamiltonian for the family of non-Abelian chiral spin
  liquids proposed recently by two of us [PRL 102, 207203 (2009)], which
  includes the Abelian chiral spin liquid proposed by Kalmeyer and Laughlin,
  as the special case $s=\frac{1}{2}$.  As we use a circular disk geometry
  with an open boundary, both the annihilation operators we identify and the
  Hamiltonians we construct from these, are exact only in the thermodynamic
  limit.
\end{abstract}



\maketitle

{\it Introduction.}---The field of two-dimensional quantum spin
liquids~\cite{anderson87s1196,kalmeyer-87prl2095,kivelson-87prb8865,wen-89prb11413,moessner-01prl1881,balents-02prb224412,greiter02jltp1029,kitaev06ap2,isakov-06prl207204,yao-07prl247203,dusuel-08prb125102,lee08s1306,greiter-09prl207203,hermele-09prl135301,balents10n199}
is witnessing a renaissance of interest in present
days~\cite{zhang-11prb075128,grover-11prl077203,yao-11prl087205,scharfenberger-11prb140404,nielsen-11jsmte11014}.
For one thing, due to advances in the computer facilities available, evidence
for spin liquid states in a range of models is
accumulating~\cite{meng-10n847,jiang-11arXiv:1112.2241}.  At the same time,
spin liquids constitute the most intricate, and in general probably least
understood, examples of topological
phases~\cite{wen89prb7387,wen04,
levin-05prb045110,chung-10prb060403,isakov-11np772},
which themselves establish another vividly studied branch of condensed matter
physics~\cite{moore10n194,hasan-10rmp3045,qi-11rmp1057}.  If a complete
description of the electronic states in the two-dimensional (2D) CuO planes of
high T$_{\rm c}$ superconductors~\cite{zaanen-06np138} ever emerges, the
theory is likely based on a spin $s=1/2$ liquid on a square lattice, which is
stabilized through the kinetic energy of itinerant holon
excitations~\cite{anderson87s1196}.

Intimately related to the field of topological phases are the concepts of
fractional quantization, and in particular fractional
statistics~\cite{wilczek90}. This field has experienced another, seemingly
unrelated renaissance of interest in recent years, due to possible
applications of states supporting excitations with non-Abelian
statistics~\cite{stern10n187} to the rapidly evolving field of quantum
computing and cryptography.  The paradigm for this class is the Pfaffian
state~\cite{moore-91npb362,greiter-92npb567}, which has been proposed to
describe the experimentally observed quantized Hall plateau at Landau level
filling fraction $\nu=\frac{5}{2}$~\cite{greiter-92npb567}.  The state
supports quasiparticle excitations which possess Majorana fermion states at
zero energy~\cite{read-00prb10267}.  Braiding of these half-vortices yields
non-trivial changes in the occupations of the Majorana fermion states, and
hence render the exchanges non-commutative or
non-Abelian~\cite{ivanov01prl268,Stern-04prb205338}.  Since this ``internal''
state vector is insensitive to local perturbations, it is preeminently suited
for applications as protected qubits in quantum
computation~\cite{kitaev03ap2,nayak-08rmp1083}.  Non-Abelian anyons are
further established in other quantum Hall states including Read-Rezayi
states~\cite{read-99prb8084}, in the non-Abelian phase of the Kitaev
model~\cite{kitaev06ap2}, the Yao--Kivelson and Yao--Lee
models~\cite{yao-07prl247203,yao-11prl087205}, and in the family of
non-Abelian chiral spin liquid (NACSL) states introduced by two of
us~\cite{greiter-09prl207203}.  Very recently, non-Abelian statistics has been
observed numerically in hard-core lattice bosons in a magnetic field, without
reference to explicit wave functions~\cite{kapit-11arXiv:1109.4561}.

In this Letter, we construct a parent Hamiltonian for the NACSL
states~\cite{greiter-09prl207203}.  These spin liquids support spinon
excitations with SU(2) level $k=2s$ statistics for spin $s$, \ie Abelian,
Ising, and Fibonacci anyons for $s=\frac{1}{2},1$, and $\frac{3}{2}$,
respectively.  The method we employ here is different from the method we used
to identify a Hamiltonian~\cite{schroeter-07prl097202,thomale-09prb104406}
which singles out the Kalmeyer--Laughlin chiral spin liquid (CSL)
state~\cite{kalmeyer-87prl2095,kalmeyer-89prb11879} as its (modulo the
two-fold topological degeneracy) unique ground state for periodic boundary
conditions (PBCs).  It is considerably simpler, applicable to the entire
family of spin $s$ NACSL states, but exact only in the thermodynamic (TD)
limit even if we impose PBCs.


{\it Chiral spin liquid states.}---The conceptually simplest way to construct
the non-Abelian chiral spin liquid (NACSL) state~\cite{greiter-09prl207203}
with spin $s$ is to combine $2s$ identical copies of Abelian CSL
states 
with spin $\frac{1}{2}$, and project the spin on each site onto spin $s$,
\newcommand{\struta}{\rule[-6pt]{0pt}{0pt}}
\newcommand{\strutb}{\rule[6pt]{0pt}{0pt}}
\begin{align*}
  \underbrace{\textstyle \bs{\frac{1}{2}}\otimes\bs{\frac{1}{2}}\otimes
    \ldots\otimes\bs{\frac{1}{2}}\struta}_{2s\strutb}
  =\bs{s}\oplus (2s-1)\cdot \bs{s\!-\!1}\oplus\ldots
\end{align*}
The projection onto the completely symmetric representation can be carried out
conveniently using Schwinger bosons~\cite{arovas-88prl531,greiter02jltp1029}.
For a circular droplet with open boundary conditions occupying $N$ sites on a
triangular or square lattice, the Abelian CSL state takes the form
\begin{align}
  \label{eq:klket}
  \ket{\psi^{\s\text{KL}}_{0}}
  &=\hspace{-6pt}\sum_{\{z_1,\ldots ,z_M\}}\hspace{-7pt}
  \psi^{\s\text{KL}}_{0}(z_1,\ldots ,z_M)\;
  {S}^+_{z_1}\cdot\ldots\cdot {S}^+_{z_M} 
  \ket{\dw\dw\ldots\dw} 
  \nonumber\\[0.2\baselineskip] 
  &=\hspace{-6pt}\sum_{\substack{\{z_1,\ldots ,z_M;\ \\[2pt]\ w_1,\ldots,w_M\}}}\hspace{-10pt}
  \psi^{\s\text{KL}}_{0}(z_1,\ldots ,z_M)\;
    {a}^+_{z_1}\ldots a^\dagger_{z_M}
    {b}^+_{w_1}\ldots b^\dagger_{w_M}%
  \vac\!
  \nonumber\\[0.2\baselineskip] 
  &\equiv \Psi^{\s\text{KL}}_{0}[a^\dagger,b^\dagger] \vac\!,
\end{align}
where
\begin{align}
  \label{eq:klpsi}
  \psi^{\s\text{KL}}_{0}[z]
  &=\prod_{i<i}^M\,(z_i-z_j)^2\,\prod_{i=1}^M\,G(z_i)\,e^{-\frac{1}{4}|z_i|^2}
\end{align}
is a bosonic quantum Hall state in the complex ``particle'' coordinates
$z_i\equiv x_i+\i y_i$ supplemented by a gauge factor $G(z_i)$,
$M=\frac{N}{2}$, $a^\dagger$ and $b^\dagger$ are Schwinger boson creation
operators~\cite{schwinger65proc,arovas-88prl531,greiter02jltp1029}, and the
$w_k$'s are those lattice sites which are not occupied by any of the $z_i$'s.
In this notation, we can write the spin $s$ state obtained by the projection
as
\begin{equation}
  \label{eq:psi0schwinger}
  \ket{\psi^{s}_0}
  =\Big(\Psi^{\s\text{KL}}_0\big[a^\dagger ,b^\dagger\big]\Big)^{2s}\vac.
\end{equation} 
The lattice may be anisotropic; we have chosen the lattice constants such that
the area of the unit cell spanned by the primitive lattice vectors is set to
$2\pi$.  For a triangular or square lattice with lattice positions given by
$\eta_{n,m}=na+mb$, where $a$ and $b$ are the primitive lattice vectors in the
complex plane and $n$ and $m$ are integers, the gauge phases are simply
$G(\eta_{n,m})= (-1)^{(n+1)(m+1)}$~\cite{zou-89prb11424,
  kalmeyer-89prb11879}.

The NACSL state can alternatively be written as
\begin{equation}
  \label{eq:psi0ket}
  \ket{\psi^s_0}\;=\sum_{\{z_1,\dots,z_{sN}\}} 
  \psi^s_0(z_1,\dots,z_{SN})\
  \tilde{S}_{z_1}^{+}\cdot\ldots\cdot\tilde{S}_{z_{sN}}^{+} 
  \ket{-s}_N,
\end{equation}
where 
$\ket{-s}_N\equiv\otimes_{\alpha=1}^N \ket{s,-s}_{\alpha}$
is the ``vacuum'' state in which all the spins are maximally polarized
in the negative $\hat z$-direction, and $\tilde{S}^{+}$ are
re-normalized spin flip operators which satisfy
\begin{equation}
  \label{eq:Stilde^n}
  \frac{1}{\sqrt{(2s)!}}(a^\dagger)^n (b^\dagger)^{(2s-n)}\vac 
  =(\tilde{S}^+)^n\ket{s,-s}.
\end{equation}
In a basis in which $S^\z$ is diagonal, we may write 
\begin{align}
  \label{eq:Stilde}
  \tilde{S}^{+} 
  = \frac{1}{s-{S}^\z+1}\, S^{+}.
\end{align}
Note that \eqref{eq:Stilde^n} implies
\begin{align}
  \label{eq:S-Stilde^n}
  S^-(\tilde{S}^+)^n\ket{s,-s}
  &=n(\tilde{S}^+)^{n-1}\ket{s,-s}.
\end{align}
The wave function for the spin $s$ state \eqref{eq:psi0schwinger} are then
effectively given by bosonic Read--Rezayi states~\cite{read-99prb8084} for
renormalized spin flips,
\begin{equation}
  \label{eq:psirr}
  \psi^s_0[z]
  =\prod_{m=1}^{2s}\!\Biggl\{
    \prod_{\substack{i,j=(m-1)M+1\\[1pt] i<j}}^{mM}\vspace{3pt}(z_i-z_j)^2 
  \Biggr\}\!
  \prod_{i=1}^{sN}G(z_i) e^{-\frac{1}{4}|z_i|^2}.
\end{equation}
which we understand to be completely symmetrized over the ``particle''
coordinates $z_i$.  For $s=1$, they take the form of a Moore--Read
state~\cite{moore-91npb362,
  greiter-92npb567}
\begin{equation}
  \label{eq:psipf}
  \psi^{s=1}_0[z]
  =\text{Pf}\left(\frac{1}{z_{i}-z_{j}}\right)\prod_{i<j}^{N}(z_i-z_j)
  \prod_{i=1}^{sN}G(z_i) e^{-\frac{1}{4}|z_i|^2}.
\end{equation}

For the considerations below, it is convenient to write the state in the
form
\begin{align}
  \label{eq:psi0prodket}
  \ket{\psi^{s}_0}
  &=\left[
    \sum_{\{z_1,\ldots ,z_M\}}
    \psi^{\s\text{KL}}_{0}(z_1,\ldots ,z_M)\;
    \tilde{S}_{z_1}^{+}\cdot\ldots\cdot\tilde{S}_{z_{M}}^{+}
  \right]^{2s}\vac.
\end{align}
Since the Abelian KL CSL $\ket{\psi^{\s\text{KL}}_{0}}$ is an exact spin
singlet in the TD limit $N\to\infty$, and is an approximate
singlet for finite $N$, the same holds for the NACSL $\ket{\psi^{s}_0}$ as
well.  This follows from the construction of the Schwinger boson projection
\eqref{eq:psi0schwinger}, but can also be verified directly using Perelomov's
identity (see \eqref{eq:perel} in the supplementary
material)~\cite{perelomov71tmp156}.  The Abelian and non-Abelian CSL states
trivially violate parity (P) and and time reversal (T) symmetry.

{\it Ground state annihilation operators.}---In the TD limit
$N\to\infty$, the NACSL ground states are annihilated by
\begin{align}
  \label{eq:Omegadef}
    \OaS &=\sum_{\substack{\beta=1\\\beta\ne\alpha}}^N 
    \frac{1}{\ea-\eb} (\Sa^-)^{2s} \Sb^-,\quad 
    \OaS \ket{\psi^s_{0}} = 0 \ \forall\, \alpha,
\end{align}
as we will verify now.

Let us consider the action of $(\Sa^-)^{2s} \Sb^-$ on 
$\ket{\psi^s_{0}}$ written in the form 
\eqref{eq:psi0prodket}.  Since $\psi^{\s\text{KL}}_{0}(z_1,\ldots ,z_M)$
vanishes whenever two arguments $z_i$ coincide, one of the $z_i$'s
in each of the $2s$ copies 
in \eqref{eq:psi0prodket} must equal $\ea$; since
$\psi^{\s\text{KL}}_{0}(z_1,\ldots ,z_M)$ is symmetric under
interchange of the $z_i$'s and we count each distinct configuration in
the sums over ${\{z_1,\ldots ,z_M\}}$ only once, we may take
$z_1=\ea$.  Regarding the action of $\Sb^-$ on
\eqref{eq:psi0prodket}, we have to distinguish between
configurations with $n=0,1,2,\ldots,2s$ re-normalized spin flips
$\tilde{S}_\b^+$ at site $\b$.  Since the state is symmetric under
interchange of the $2s$ copies
, we may assume that the $n$ spin flips are present in the first $n$
copies, and account for the restriction through ordering by a
combinatorial factor.  This yields
\begin{widetext}
  \begin{align}
    \hspace{2pt}&\hspace{-2pt}
    (\Sa^-)^{2s} \Sb^-\ket{\psi^{s}_0}
    =(\Sa^-)^{2s} \Sb^-\sum_{n=0}^{2s}\binom{2s}{n}\hspace{-3pt}
    \left[\sum_{\{z_3,\ldots ,z_M\}}
      \psi^{\s\text{KL}}_{0}(\ea,\eb,z_3,\ldots)\,
      \tilde{S}_{\a}^{+}\tilde{S}_{\b}^{+}
      \tilde{S}_{z_3}^{+}\cdot\ldots\cdot\tilde{S}_{z_{M}}^{+} \right]^n
    \nonumber\\*[0.2\baselineskip]
    &\hspace{151pt}\cdot\left[\sum_{\substack{\{z_2,\ldots ,z_M\}\ne\eb}}
      \psi^{\s\text{KL}}_{0}(\ea,z_2,\ldots )\,
      \tilde{S}_{\a}^{+}\tilde{S}_{z_2}^{+}\cdot\ldots\cdot\tilde{S}_{z_{M}}^{+}
    \right]^{2s-n}\!\vac
    \nonumber\\[0.2\baselineskip]
    &=(2s)!\, 2s\!  \left[\sum_{\{z_2,\ldots
        ,z_M\}}\psi^{\s\text{KL}}_{0}(\ea,\eb,z_3,\ldots ,z_M)
      \,\tilde{S}_{z_3}^{+}\cdot\ldots\cdot\tilde{S}_{z_{M}}^{+}\right]
    \sum_{n=1}^{2s}\binom{2s-1}{n-1} 
    \nonumber\\*[0.2\baselineskip]
    &\quad\cdot \left[\sum_{\{z_3,\ldots
        ,z_M\}} \psi^{\s\text{KL}}_{0}(\ea,\eb,z_3,\ldots ,z_M)\,
      \tilde{S}_{\b}^{+}\tilde{S}_{z_3}^{+}\cdot\ldots\cdot\tilde{S}_{z_{M}}^{+}
    \right]^{n-1}
    \left[\sum_{\substack{\{z_2,\ldots
          ,z_M\}\ne\eb}} \psi^{\s\text{KL}}_{0}(\ea,z_2,\ldots ,z_M)\,
      \tilde{S}_{z_2}^{+}\cdot\ldots\cdot\tilde{S}_{z_{M}}^{+}
    \right]^{2s-n}\vac
    \nonumber\\[0.2\baselineskip]
    &=(2s)!\, 2s\!  \left[\sum_{\{z_3,\ldots
        ,z_M\}}\psi^{\s\text{KL}}_{0}(\ea,\eb,z_3,\ldots ,z_M)
      \,\tilde{S}_{z_3}^{+}\cdot\ldots\cdot\tilde{S}_{z_{M}}^{+}\right]
    \cdot\left[\sum_{\substack{\{z_2,\ldots ,z_M\}}}
      \psi^{\s\text{KL}}_{0}(\ea,z_2,\ldots ,z_M)\,
      \tilde{S}_{z_2}^{+}\cdot\ldots\cdot\tilde{S}_{z_{M}}^{+}
    \right]^{2s-1}\vac ,\nonumber
  \end{align}
  where we have used \eqref{eq:S-Stilde^n}.  This implies
  \renewcommand{\struta}{\rule[-18pt]{0pt}{0pt}}
  \begin{align}
    \OaS \ket{\psi^{s}_0} 
    &=(2s)!\, 2s\!\left[\sum_{\{z_3,\ldots ,z_M\}}\right.
    \underbrace{\sum_{\beta=1}^N
      \frac{\psi^{\s\text{KL}}_{0}(\ea,\eb,z_3,\ldots
        ,z_M)}{\ea-\eb}\struta}_{=0}
    \,\tilde{S}_{z_3}^{+}\cdot\ldots\cdot\tilde{S}_{z_{M}}^{+}
    \left]\phantom{\sum_{\{z_3,\ldots ,z_M\}}} \right.\hspace{-60pt}
    \nonumber\\* 
    &\phantom{=(2s)!\, 2s\!}\hspace{-2.7pt}
    \cdot
    \left[\sum_{\substack{\{z_2,\ldots ,z_M\}}}
      \psi^{\s\text{KL}}_{0}(\ea,z_2,\ldots ,z_M)\,
      \tilde{S}_{z_2}^{+}\cdot\ldots\cdot\tilde{S}_{z_{M}}^{+}
    \right]^{2s-1}\vac=0,\nonumber
  \end{align}
%
\end{widetext}
where we have used the Perelomov identity~\cite{perelomov71tmp156}, which
states that any infinite lattice sum of $e^{-\frac{1}{4}|\eb|^2} G(\eb)$
times any analytic function of $\eb$ vanishes.  (Strictly speaking,
Perelomov~\cite{perelomov71tmp156} only considered a square lattice.  The
identity, however, holds for any 2D lattice with a single site per unit cell,
as we show in the supplementary material.)

\def\prefac{\frac{1}{2}}
\def\prefac{}

{\it Parent Hamiltonian.}---A Hermitian, positive semi-definite, and
translationally invariant operator which annihilates $\ket{\psi^{s}_0}$ is
given by
\begin{align}
  \label{eq:h0}
  \Gamma&\equiv\prefac\sum_{\a=1}^N {\OaS}^\dagger\OaS
  =\prefac\sum_{\substack{\a,\b,\c\\ \a\ne \b,\c}}
    \omega_{\a\b\c}(\Sa^+)^{2s}(\Sa^-)^{2s}\Sb^+\Sc^-,
\end{align}
where 
\begin{align}
  \label{eq:omegaabc}
  \omega_{\a\b\c}\equiv\frac{1}{\eab-\ebb}\frac{1}{\ea-\ec}.
\end{align}
This operator is not invariant under SU(2) spin rotations, but rather consists
of a scalar, vector, and higher tensor components up to order $4s+2$.  Since
the NACSL states
$\ket{\psi^{s}_{0}}$ 
are spin singlets, and are annihilated by $\Gamma$, all these tensor components
must annihilate the state individually.  The scalar component of $\Gamma$, which
we denote as $\{\Gamma\}_0$, provides us with an SU(2) spin rotationally
invariant parent Hamiltonian.

To obtain the projected operator $\{\Gamma\}_0$, we follow the method described
in detail in ref.~\cite{Greiter11}, and summarize here only the most important
steps.  With the tensor content of $\Sb^+\Sc^-$ given by
\begin{align}
  \label{eq:Sb+Sc-}
  \Sb^+\Sc^-&=\frac{2}{3}\bSb\bSc - \text{i}(\bSb\times\bSc)^\z
    -\frac{1}{\sqrt{6}}\,T_{\b\c}^0,
\end{align}
where 
\begin{align}
  \label{eq:Tbc}
  T_{\b\c}^0
  &=\frac{2}{\sqrt{6}}\big(3 \Sb^\z\Sc^\z - \bSb\bSc\big)
\end{align}
is the $m=0$ component of the second order tensor, we only need to know the
scalar, vector and 2nd order tensor components of $(\Sa^+)^{2s}(\Sa^-)^{2s}$
in order to obtain the scalar component of $\Gamma$.  These are given 
by (see Sec.\ 5.3.2 of~\cite{Greiter11})
\begin{align}
  \label{eq:c:S+^2sS-^2s}
  (\Sa^+)^{2s}(\Sa^-)^{2s}
  &=a_0\,\Big\{ 1 + a\, \Sa^\z + b\,T_{\a\a}^0  + \text{higher orders} \Big\} 
\end{align}
where 
\begin{align}
  \label{eq:c:a_0abd} 
  a_0=\frac{{(2s)!}^2}{2s+1},\
  a=\frac{3}{s+1},\
  b=\frac{\sqrt{6}}{2}\frac{5}{(s+1)(2s+3)}.
\end{align}
The scalar component of $\Gamma$ is hence given by
\begin{align}
  \label{eq:Gamma0}
  \big\{\Gamma\big\}_0
  &=a_0\sum_{\substack{\a,\b,\c\\ \a\ne\b,\c}}\omega_{\a\b\c} 
  \nonumber\\[0.2\baselineskip] 
  &\cdot\left[\frac{2}{3}\bSb\bSc -\frac{\i a}{3}\bSa(\bSb\times\bSc)
    - \frac{b}{\sqrt{6}}\left\{T_{\a\a}^0 T_{\b\c}^0\right\}_0\right].
\end{align}
With $\bSb\times\bSb=\i\bSb$ and (see Sec.\ 4.5.3 of~\cite{Greiter11})  
\begin{align}
  \label{eq:5TaaTbc_0}
  5\,\big\{T_{\a\a}^0T_{\b\c}^0 \big\}_0 
  \hspace{0pt}&\hspace{0pt}
  = -\frac{4}{3} \bSa^2(\bSb\bSc) + 2\delta_{\b\c}\bSa\bSb
  \nonumber\\[0.2\baselineskip] 
  &\quad + 2\bigl[(\bSa\bSb)(\bSa\bSc)+ (\bSa\bSc)(\bSa\bSb)\bigr],
\end{align}
we obtain the final parent Hamiltonian
\begin{align}
  \label{eq:h}
  &H^s=\sum_{\substack{\a\ne\b}}\omega_{\a\b\b} 
  \bigg[s(s+1)^2+\bSa\bSb-\frac{(\bSa\bSb)^2}{(s+1)}\bigg]
  \nonumber\\[0.2\baselineskip] 
  &+\hspace{0pt}\sum_{\substack{\a,\b,\c\\ \a\ne\b\ne\c\ne\a}}\hspace{-7pt}
  \omega_{\a\b\c}
  \bigg[(s+1)\bSb\bSc - \frac{2s+3}{2(s+1)}\i\bSa(\bSb\times\bSc)
  \nonumber\\[-0.2\baselineskip] 
  &\hspace{75pt}
  -\frac{(\bSa\bSb)(\bSa\bSc)+(\bSa\bSc)(\bSa\bSb)}{2(s+1)}\bigg]. 
\end{align}
(It is related to \eqref{eq:Gamma0} via
$\big\{\Gamma\big\}_0=2a_0/(2s+3)\, H^s$.)  
This Hamiltonian is approximately valid for any finite disk with $N$ lattice
sites, and becomes exact in the TD limit $N\to\infty$, where
$H^s\ket{\psi^s_{0}} = 0$.
Note that the $\bSa(\bSb\times\bSc)$ term explicitly breaks P and T.  (It
would be highly desirable to identify a parent Hamiltonian which is P and T
invariant, such that the ground states violate these symmetries spontaneously,
but we have so far not succeeded in finding one.)

{\it The special case $s=\frac{1}{2}$.}---Since ${\Sa^+}^2=0$ for $s=\frac{1}{2}$,
$T_{\a\a}^m=0$ for all $m$, and $\big\{T_{\a\a}^0T_{\b\c}^0 \big\}_0=0$.  This
simplifies \eqref{eq:Gamma0} significantly, and yields the parent Hamiltonian
\begin{align}
  \label{eq:hhalf}
  &H^{s=\frac{1}{2}} 
  =\sum_{\substack{\a\ne\b}}\omega_{\a\b\b} 
  \bigg[\frac{3}{4}+\bSa\bSb\bigg]
  \nonumber\\[0.2\baselineskip] 
  &+\hspace{-3pt}\sum_{\substack{\a,\b,\c\\ \a\ne\b\ne\c\ne\a}}\hspace{-3pt}
  \omega_{\a\b\c}
  \big[\bSb\bSc -\i \bSa(\bSb\times\bSc)\big]
\end{align}
(It is related to \eqref{eq:Gamma0} via
$\big\{\Gamma\big\}_0={2a_0}/{3}\,H^{s=\frac{1}{2}}$.)  In contrast to the earlier
parent Hamiltonian proposed in
ref.~\cite{schroeter-07prl097202,thomale-09prb104406} (SKTG) for the
Abelian KL CSL \eqref{eq:klpsi} with periodic boundary conditions, 
\eqref{eq:hhalf} is not exact for finite $N$.  It is considerably simpler
then the SKTG model, and, like \eqref{eq:h}, becomes exact in the TD limit.

{\it Remarks on periodic boundary conditions.}---It is rather straightforward
to formulate the model on a torus.  For simplicity, we choose the lattice
constant $a$ real, and
$b$ such that the imaginary part $\Im(b)>0$.  We implement PBCs in both
directions by identifying the sites $z_i$, $z_i + L$, and $z_i +L\tau $, where
$L = n_1 a$, $L\tau = n_\tau a+ m_\tau b$, and $\Im(\tau)>0$.  $n_1$ and
$m_\tau$ are positive integers such that the number of sites $N=n_1 m_\tau$ is
even, and $n_\tau$ is an integer.  We place the lattice sites at positions
\begin{align}
  \eta_{n,m}=\left(n-\frac{n_1-1}{2}\right) a +\left(m-\frac{m_\tau-1}{2}\right) b,
\end{align}
with $n=0,1,\ldots,n_1\hspace{-2pt}-\hspace{-2pt}1$ and
$m=0,1,\ldots,m_\tau\hspace{-2pt}-\hspace{-2pt}1$.  
Then the wave function of the NACSL \eqref{eq:psirr} takes the
form 
\begin{align}
  \label{eq:psirr.torus}
  \psi^s_0[z]
  =&\prod_{m=1}^{2s}\!\Biggl\{
   \prod_{\substack{i,j=(m-1)M+1\\[1pt] i<j}}^{mM} 
    \vartheta_{\frac{1}{2}\hspace{-1pt},\hspace{-1pt}\frac{1}{2}}
    \bigl(\textstyle{\frac{1}{L}}(z_i-z_j)\bigr|\tau\bigl)^2\bigr.
  \Biggr.
  \nonumber\\ 
   &\hspace{6pt}\cdot\prod_{\nu=1}^2 
    \vartheta_{\frac{1}{2}\hspace{-1pt},\hspace{-1pt}\frac{1}{2}}
    \bigl(\textstyle{\frac{1}{L}}(Z_m-Z_{\nu,m})\bigr|\tau\bigl)\bigr.
    {\Biggl\}\cdot \displaystyle \prod_{i=1}^{sN}G(z_i) e^{-\frac{1}{2}y_i^2}, \Biggr.}
\end{align}
where $\vartheta_{\frac{1}{2}\hspace{-1pt},\hspace{-1pt}\frac{1}{2}}(z|\tau)$
is the odd Jacobi theta function~\cite{mumford83}, and
\begin{align}
  Z_m\equiv\sum_{i=(m-1)M+1}^{mM}z_i,\hspace{10pt}
  Z_{1,m}=-Z_{2,m}, 
\end{align}
are the center-of-mass coordinates and zeros, respectively.
The latter can be chosen anywhere within the principal region
bounded by the four points $\frac{1}{2}(\pm{n_1}{a}\pm{m_\tau}{b})$,
and encode the $(2s+1)$-fold topological degeneracy of the
NACSL~\cite{scharfenberger-11prb140404}.  The gauge factor in
\eqref{eq:psirr.torus} is given by
\begin{align}
  G(\eta_{n,m})=(-1)^{m_\tau n+m}e^{-\i\pi\frac{\Re(b)}{a}m(m_\tau-1-m)},
\end{align}
where $\Re(b)$ is the real part of $b$.
 
The NACSL \eqref{eq:psirr.torus} is approximately annihilated by
\begin{align}
  \label{eq:Omegadef.torus}
  \OaS &=\sum_{\substack{\beta=1\\\beta\ne\alpha}}^N 
  \frac{\vartheta_{u,v}
    \bigl(\textstyle{\frac{1}{L}}(\ea-\eb)\bigr|\tau\bigl)\bigr.}
       {\vartheta_{\frac{1}{2}\hspace{-1pt},\hspace{-1pt}\frac{1}{2}}
    \bigl(\textstyle{\frac{1}{L}}(\ea-\eb)\bigr|\tau\bigl)\bigr.}
    (\Sa^-)^{2s} \Sb^-
\end{align}
for all $\a$,
where we can choose any of the three even Jacobi theta functions
in the numerator: $(u,v)$=$(0,0)$, $(0,\frac{1}{2})$, or $(\frac{1}{2},0)$.
Note that $\OaS \ket{\psi^s_0}$
is not strictly periodic, but only quasiperiodic, due
to the shift of the boundary phases inherent in \eqref{eq:Omegadef.torus}.
The statement $\OaS \ket{\psi^s_0}\approx 0$ becomes exact as $N\to\infty$.

The NACSL \eqref{eq:psirr.torus} is hence the approximate ground state of
\eqref{eq:h} (and for $s=\frac{1}{2}$
also of \eqref{eq:hhalf}) with \eqref{eq:omegaabc} replaced by
\begin{align}
  \label{eq:omegaabc.torus}
  \omega_{\a\b\c}=\left(\frac{\vartheta_{u,v}
    \bigl(\textstyle{\frac{1}{L}}(\ea-\eb)\bigr|\tau\bigl)\bigr.}
  {\vartheta_{\frac{1}{2}\hspace{-1pt},\hspace{-1pt}\frac{1}{2}}
    \bigl(\textstyle{\frac{1}{L}}(\ea-\eb)\bigr|\tau\bigl)\bigr.}
    \right)^{\hspace{-4pt}*}
  \frac{\vartheta_{u,v}
    \bigl(\textstyle{\frac{1}{L}}(\ea-\ec)\bigr|\tau\bigl)\bigr.}
  {\vartheta_{\frac{1}{2}\hspace{-1pt},\hspace{-1pt}\frac{1}{2}}
    \bigl(\textstyle{\frac{1}{L}}(\ea-\ec)\bigr|\tau\bigl)\bigr.},
\end{align}
where $*$ denotes complex conjugation.  As in the case with open boundary
conditions, the model becomes exact in the TD limit.

{\it Conclusion.}---We have identified a parent Hamiltonian for the
non-Abelian CSL states~\cite{greiter-09prl207203}, which becomes exact in the
TD limit.  This Hamiltonian should allow us to study the spinon and holon
excitations including the non-Abelian braiding properties within a concise
framework.  The construction also extends to the Abelian $s=\frac{1}{2}$
Kalmeyer--Laughlin CSL~\cite{kalmeyer-87prl2095,kalmeyer-89prb11879}, where it
is likewise exact only as the number of sites $N\to\infty$, but is
considerably simpler that the 
SKTG Hamiltonian~\cite{schroeter-07prl097202,thomale-09prb104406}. 

{\it Acknowledgments.}---MG is supported by the German Research Foundation
under grant FOR 960.  RT is supported by an SITP fellowship at Stanford
University.

{\it Note added.}---After this work was completed, we became aware of a
manuscript by Nielsen, Cirac, and Sierra~\cite{nielsen-1201.3096}, in which
they derive the $s=\frac{1}{2}$ Hamiltonian \eqref{eq:hhalf} using null
operators in the conformal correlators of the SU(2) level $k=1$ 
Wess--Zumino--Witten model.



\newpage

\section*{Supplementary material}

In this supplement, we proof the Perelomov identity~\cite{perelomov71tmp156} for arbitrary 2D lattices
using Fourier transformation.

\emph{The Perelomov identity}.---Consider a
lattice 
spanned by $\eta_{n,m}=na+mb$ in the complex plane, with $n$ and $m$ integer
and the area of the unit cell $\Omega$ spanned by the primitive lattice
vectors $a$ and $b$ set to $2\pi$,
\begin{equation}
  \label{eq:area}
  \Omega=\left|\Im(a\bar b)\right|
  = 2\pi
\end{equation}
where $\Im$ denotes the imaginary part.  Let $G(\eta_{n,m})=
(-1)^{(n+1)(m+1)}$.  Then
\begin{equation}
  \label{eq:perel}
  \sum_{n,m} P(\eta_{n,m}) G(\eta_{n,m}) e^{-\frac{1}{4}|\eta_{n,m}|^2}=0
\end{equation}
for any polynomial $P$ of $\eta_{n,m}$.

\vspace{3pt}
\emph{Proof.}---It is sufficient to proof the identity for the
generating functional
\begin{equation}
  \label{eq:genfun}
  \sum_{n,m} e^{\frac{1}{2}\eta_{n,m}\bar z} G(\eta_{n,m}) e^{-\frac{1}{4}|\eta_{n,m}|^2}=0.
\end{equation}
Since $G(\eta_{n,m})$ takes the value $-1$ on a lattice with twice the original
lattice constants, we may rewrite this as
\begin{equation}
  \label{eq:genfun1}
  \sum_{n,m} e^{\frac{1}{2}\eta_{n,m} \bar z} e^{-\frac{1}{4}|\eta_{n,m}|^2}-
  2\sum_{n,m} e^{\eta_{n,m} \bar z} e^{-|\eta_{n,m}|^2}=0.
\end{equation}
Kalmeyer and Laughlin~\cite{kalmeyer-89prb11879} observed that for the square
lattice, the second sum in \eqref{eq:genfun1} can be expressed as a sum of the
Fourier transform of the function we sum over in the first term.  We
demonstrate here that their proof can be extended to arbitrary lattices.

To begin with, we define the Fourier transform in complex coordinates
\begin{equation}
  \label{eq:ft}
  \tilde f(\zeta)=\int d^2\eta f(\eta) e^{i\Re(\eta\bar\zeta)},
\end{equation}
where $\Re$ denotes the real part and we have used \eqref{eq:area}.  Since the
area of the unit cell of our lattice is taken to be $2\pi$, the reciprocal
lattice is given by the original lattice rotated by $\frac{\pi}{2}$ in the
plane without any rescaling of the lattice constants.  In complex coordinates,
\begin{equation}
  \label{eq:reciprocal}
  \zeta_{n',m'}=i(n'a+m'b),
\end{equation}
as this immediately implies
\begin{eqnarray}
  \label{eq:reciprocalcheck}
  \bs{R}_{n,m}\cdot \bs{K}_{n',m'}\nonumber
  &=&\Re(\eta_{n,m}\bar\zeta_{n',m'})=\\\nonumber
  &=&\Re\left((na+mb)(-i)(n'\bar a+m'\bar b)\right)\\\nonumber
  &=&n m'\Im(a\bar b)+m n'\Im(b\bar a)\\\nonumber
  &=&2\pi\cdot\text{integer}.
\end{eqnarray}
Then 
\begin{equation}
  \label{eq:sumsum}
 \sum_{n',m'}\tilde f(\zeta_{n',m'}) = \Omega\sum_{n,m} f(\eta_{n,m}).
\end{equation}
Eq.\ \eqref{eq:sumsum} follows directly from
\begin{equation}
  \label{eq:delta}
  \sum_{n',m'} e^{i\Re(\eta\bar\zeta_{n',m'})}
    = \Omega\sum_{n,m} \delta^{(2)}(\eta_{n,m}-\eta),
\end{equation}
which is just the 2D equivalent of the (Dirac comb) identity
\begin{equation}
  \label{eq:delta1d}
  \sum_{n'=-\infty}^\infty e^{2\pi i n' x}
    = \sum_{n=-\infty}^\infty  \delta (x-n)
\end{equation}
The r.h.s.\ of \eqref{eq:delta1d} is obviously zero if $x$ is not an
integer, and manifestly periodic in x with period 1.  To verify
the normalization, observe that since for any $N$ odd,
\begin{equation}
  \label{eq:simplesum}\nonumber
  \sum_{n'=-\frac{N-1}{2}}^{+\frac{N-1}{2}}e^{2\pi i n' y/N}
  =\begin{cases}
    N &\text{for}\ y=N\cdot\text{integer}\\[4pt]
    \,0 &\text{otherwise}.
  \end{cases}
\end{equation}
This implies
\begin{equation}
  \label{eq:simplesum2}\nonumber
  \frac{1}{N}\sum_{y=-\frac{N-1}{2}}^{+\frac{N-1}{2}}
  \sum_{n'=-\frac{N-1}{2}}^{+\frac{N-1}{2}}e^{2\pi i n' y/N}=1,
\end{equation}
which in the limit $N\rightarrow\infty$ is equivalent to
\begin{equation}
  \label{eq:simplesum3}\nonumber
  \int_{-\frac{N}{2}}^{+\frac{N}{2}}\frac{dy}{N}
  \sum_{n'=-\frac{N-1}{2}}^{+\frac{N-1}{2}}e^{2\pi i n' y/N}=1
\end{equation}
Substituting $x=y/N$ yields
\begin{equation}
  \label{eq:simplesum4}\nonumber
  \int_{-\frac{1}{2}}^{+\frac{1}{2}}dx
  \sum_{n'=-\infty}^\infty e^{2\pi i n' x}=1,
\end{equation}
which proves the normalization in \eqref{eq:delta1d}. 

We proceed by evaluation of the Fourier transform of 
$f(\eta)=e^{\frac{1}{2}\eta \bar z} e^{-\frac{1}{4}|\eta|^2}$:
\begin{eqnarray}
  \label{eq:FTofeta}\nonumber
  \tilde f(\zeta)
  &=&\int d^2\eta\, e^{\frac{1}{2}\eta \bar z} e^{-\frac{1}{4}|\eta|^2} 
  e^{i\Re(\eta\bar\zeta)}\\\nonumber
  &=&\int d^2\eta\, e^{\frac{1}{2}\eta \bar z} e^{-\frac{1}{4}|\eta|^2} 
  e^{\frac{i}{2}(\eta\bar\zeta+\bar\eta\zeta)}\\ 
  &=&4\pi e^{-|\zeta|^2 +i\zeta\bar z}
\end{eqnarray}
where we have used the integral
\begin{eqnarray}
  \label{eq:gaussintegral}\nonumber
  \int d^2\eta\!&F(\eta)\!&  e^{-\frac{1}{\alpha}(|\eta|^2- \bar\eta w)}
  \\\nonumber
  &=&\!F(\alpha \partial_{\bar w})
  \int d^2\eta\, e^{-\frac{1}{\alpha}(|\eta|^2- \bar\eta w-\eta\bar w)}
  \biggl|_{\bar w=0}\biggr.  \\\nonumber
  &=&\!F(\alpha \partial_{\bar w})
  \int d^2\eta\, e^{-\frac{1}{\alpha}(|\eta-w|^2-w\bar w)} 
  \biggl|_{\bar w=0}\biggr.\\\nonumber
  &=&\!\alpha\pi\, F(\alpha \partial_{\bar w})
  e^{\frac{1}{\alpha}w\bar w}\;=\;\alpha\pi\, F(w) \biggl|_{\bar w=0}\biggr.
\end{eqnarray}
with $F(\eta)=e^{\frac{1}{2}\eta\bar z + \frac{i}{2}\eta\bar\zeta}$,
$\alpha=4$, and $w=2i\zeta$.

Substituting \eqref{eq:FTofeta} into \eqref{eq:sumsum} we obtain
\begin{eqnarray}
  \label{eq:substituted}
  \sum_{n,m} f(\eta_{n,m}) = 2\sum_{n',m'}  e^{-|\zeta_{n',m'}|^2 +i\zeta_{n',m'}\bar z}
\end{eqnarray}
If we now substitute $n'= -n$, $m'= -m$, and hence
$i\zeta_{n',m'}=\eta_{n,m}$ into the r.h.s.\ of \eqref{eq:substituted},
we obtain \eqref{eq:genfun1}.  This completes the proof.


\vfill\eject
\end{document}